\PassOptionsToPackage{table}{xcolor}
\documentclass[letterpaper]{article} %
\usepackage{aaai2027}  %
\usepackage[hyphens]{url}  %
\usepackage{graphicx} %
\usepackage{natbib}  %
\usepackage{caption} %
\usepackage{algorithm}
\usepackage{algorithmic}
\usepackage{newfloat}
\usepackage{listings}

\DeclareCaptionStyle{ruled}{labelfont=normalfont,labelsep=colon,strut=off} %
\floatstyle{ruled}
\newfloat{listing}{tb}{lst}{}
\floatname{listing}{Listing}

\usepackage{booktabs}
\usepackage{amsmath}
\usepackage{amssymb}
\usepackage{makecell}
\usepackage{multirow}
\usepackage{subcaption}

\usepackage{tikz}

\usepackage{booktabs}  %
\usepackage{multirow}  %
\usepackage{graphicx}  %

\title{CoQui: A Coordinate-Conditioned Quantum Implicit Generative Adversarial Network for End-to-End Image Generation}
\author{
    Xue Yang\textsuperscript{\rm 1,\rm 2,\rm 3,\rm 4},
    Rigui Zhou\textsuperscript{\rm 1,\rm 2}\corresponding,
    ShiZheng Jia\textsuperscript{\rm 1,\rm 2},
    Dax Enshan Koh\textsuperscript{\rm 3,\rm 5}\corresponding,
    Siong Thye Goh\textsuperscript{\rm 4,\rm 6}\corresponding,
    Young-Wook Cho\textsuperscript{\rm 3},
    YaoChong Li\textsuperscript{\rm 1,\rm 2},
    Xuezhi Ma\textsuperscript{\rm 3},
    Hongyu Chen\textsuperscript{\rm 7},
    Xin Wang\textsuperscript{\rm 8}
}
\affiliations{
    \textsuperscript{\rm 1}School of Information Engineering, Shanghai Maritime University, Shanghai 201306, China\\
    \textsuperscript{\rm 2}Research Center of Intelligent Information Processing and Quantum Intelligent Computing, Shanghai 201306, China\\
    \textsuperscript{\rm 3}Quantum Innovation Centre (Q.InC), Agency for Science, Technology and Research (A*STAR), 2 Fusionopolis Way, Innovis, \#08-03, Singapore 138634, Republic of Singapore\\
    \textsuperscript{\rm 4}Institute of Advanced Intelligence and Computing (IAIC), Agency for Science, Technology and Research (A*STAR), 1 Fusionopolis Way, \#16-16 Connexis, Singapore 138632, Republic of Singapore\\
    \textsuperscript{\rm 5}Engineering Cluster, Singapore Institute of Technology, 1 Punggol Coast Road, Singapore 828608, Republic of Singapore\\
    \textsuperscript{\rm 6}Lee Kong Chian School of Business, Singapore Management University (SMU), Singapore 178899, Republic of Singapore\\
    \textsuperscript{\rm 7}School of Computer Science and Technology, Tongji University, Shanghai 201804, China\\
    \textsuperscript{\rm 8}Department of Automation, Tsinghua University, Beijing 100084, P. R. China\\
    rgzhou@shmtu.edu.cn,  dax.koh@singaporetech.edu.sg, goh\_siong\_thye@a-star.edu.sg
}

\begin{document}

\maketitle
\raggedbottom

\begin{abstract}
Quantum generative adversarial networks (QGANs) have recently attracted increasing attention for image generation, with the goal of using parameterized quantum circuits to model and generate image distributions. Existing approaches commonly follow a generation paradigm in which quantum-state amplitudes are mapped to pixel intensities. However, this paradigm faces two key limitations.
First, existing methods typically encode pixel locations using computational-basis indices or address qubits, causing the required quantum-state dimension or number of address qubits to grow with image resolution.
Second, existing methods jointly decode a large number of pixels from one or a few normalized quantum states. As a result, pixels compete for probability mass, making it difficult to control individual pixel values precisely.
To address these limitations, we reformulate quantum image generation as coordinate-conditioned implicit function learning. The proposed method takes spatial coordinates and latent variables as inputs, uses a classical embedding network to generate quantum circuit parameters, and evaluates a variational quantum circuit at each coordinate location. Pixel intensities are directly read out from the expectation value of a dedicated color qubit, and a complete image is obtained by querying all spatial coordinates. 
This design decouples image resolution from the number of address qubits. Furthermore, our method reads out individual pixel values from separate coordinate-conditioned quantum-state evaluations, so different pixels are not required to satisfy a shared probability-normalization constraint, enabling more flexible pixel-wise modeling.
To better support this coordinate-conditioned generation paradigm, we further design a specialized variational quantum circuit that introduces effective structural inductive bias for image generation. Simulated experiments on two benchmark datasets show that our method achieves better visual quality and quantitative performance than FRQI-based generation and PQWGAN baselines while using fewer qubits. Moreover, compared with the corresponding classical baseline, the proposed model demonstrates superior generation quality and qubit efficiency.

\end{abstract}

\definecolor{headergray}{RGB}{225,225,225}

\begin{figure*}[t]
    \centering
    \includegraphics[width=0.85\linewidth]{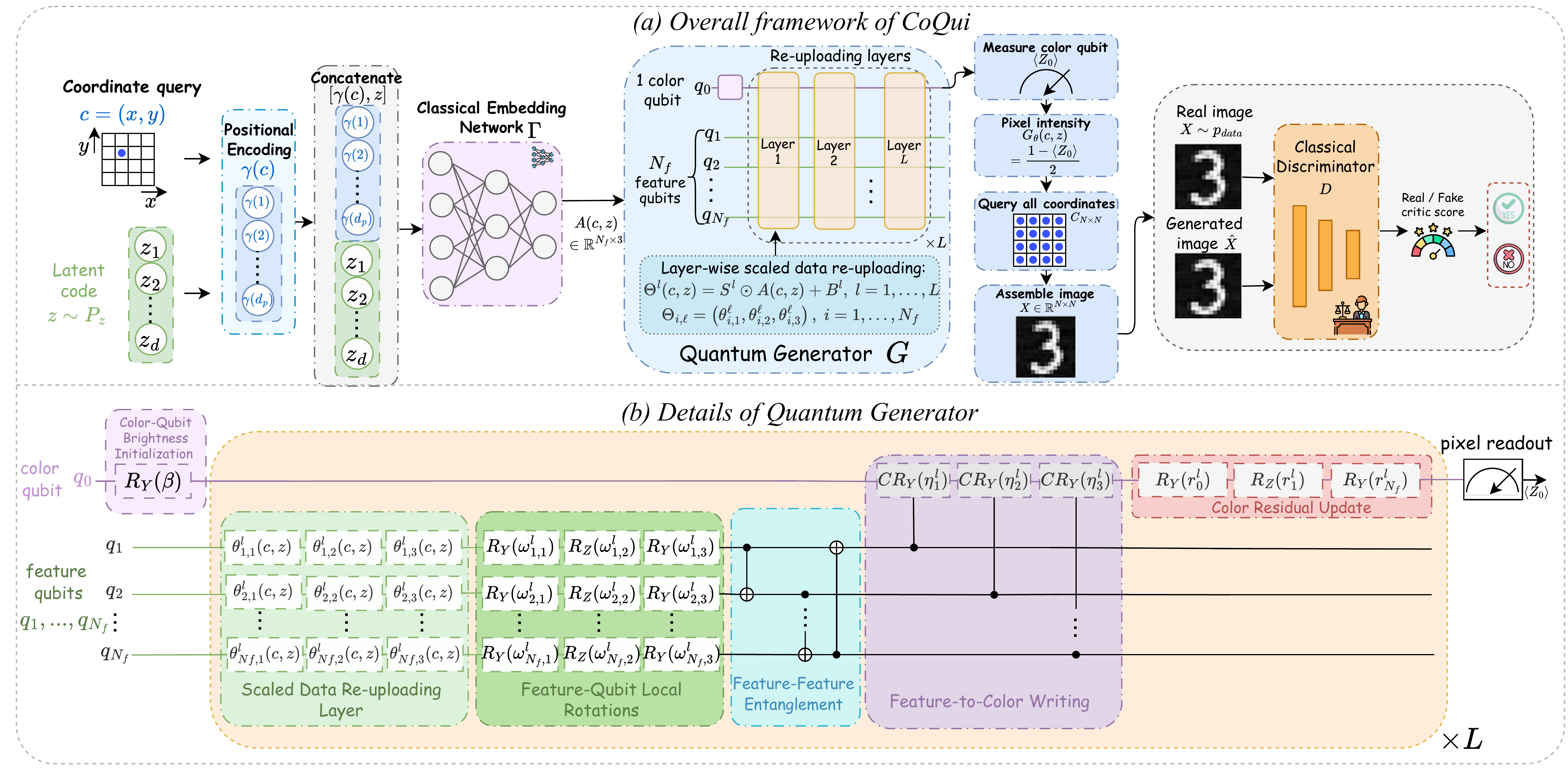}
  \caption{\textbf{Overview of the CoQui framework.} 
    \textbf{(a) Overall architecture:} Given a spatial coordinate $c=(x, y)$ and a latent code $z$, CoQui utilizes a classical embedding network $\Gamma$ to generate coordinate- and latent-conditioned parameters $A(c, z) \in \mathbb{R}^{N_f \times 3}$. These are transformed into layer-wise scaled parameters $\Theta_{i, \ell}$ for the quantum generator $G$. The pixel intensity is computed as $G(c, z) = \frac{1-\langle Z_0 \rangle}{2}$ by measuring the Pauli-$Z$ expectation of the color qubit. An entire image $\hat{X}$ is synthesized by querying all coordinates. 
    \textbf{(b) Detailed quantum generator structure:} The circuit comprises one color qubit $q_0$ and $N_f$ feature qubits. It first applies color-qubit brightness initialization to $q_0$.
	Each of the $L$ repeating layers consists of: (i) scaled data re-uploading, (ii) local feature-qubit rotations, (iii) feature-feature entanglement using a ring topology, (iv) feature-to-color writing via controlled-$R_Y$ gates, and (v) residual color-qubit updates. Finally, $q_0$ gives the pixel readout.}%
    \label{fig:CoQui_framework}
\end{figure*}

\section{Introduction}

Recent studies have shown that QGANs have attracted increasing attention and achieved promising results in image generation tasks. Existing quantum generative adversarial network (QGAN) methods for image generation commonly adopt a generation paradigm in which pixel intensities are mapped from quantum-state amplitudes.

As one of the earliest explorations in QGAN-based image generation, Huang et al. utilized quantum-state amplitudes to represent pixel values and constructed low-resolution images through a patch-based generation strategy (Huang et al. 2021). PQWGAN follows this strategy and further extends it to the high-resolution image generation setting \cite{tsang2023hybrid}. 
Although patch-based generation provides a practical way to mitigate the scalability challenge, it generates an image through multiple independently produced patches rather than modeling the entire image within a single quantum circuit.
Jäger et al. recently proposed a quantum Wasserstein GAN with spatial structural inductive bias for end-to-end high-resolution image generation \cite{jager2026scalingquantummachinelearning}. By encoding pixel positions with address qubits and pixel values with color qubits, the model generates the entire image using a single quantum circuit.

The generation paradigm based on mapping quantum-state amplitudes to pixel intensities has two major limitations. \emph{First}, under this paradigm, the number of required qubits increases together with image resolution. 
\emph{Second}, existing methods jointly recover multiple pixel values from one or a few normalized quantum states. As a result, pixels compete
for probability mass, making it difficult to control individual
pixel values precisely. 
In PQWGAN, this issue manifests as generated images whose overall brightness is lower than that of the real data distribution. It can also help explain the uneven address-amplitude distribution observed in the model of Jäger et al.

In addition, many QGAN methods that we will discuss in \textit{Related Work} use the quantum circuit only to model a classical compressed representation of the image, with the quantum circuit not directly responsible for pixel-level generation. In contrast, this work focuses on a paradigm in which the quantum circuit directly undertakes pixel-level image generation.

To address the above issues, we propose CoQui, a \textbf{co}ordinate-conditioned \textbf{qu}antum \textbf{i}mplicit GAN. The proposed framework leverages implicit neural representations (INRs) to learn a continuous mapping from spatial coordinates to signal values, thereby decoupling the number of qubits from image resolution. Moreover, unlike existing methods that jointly decode a large number of pixels from the amplitudes of one or a few quantum states, our method separately evaluates a coordinate-conditioned quantum state at each spatial coordinate and reads out the corresponding pixel value, thereby avoiding competition for probability mass among pixels.
Specifically, the model takes spatial coordinates and latent variables as inputs, first maps them into the encoding parameters of a parameterized quantum circuit (PQC) through a classical network, and then directly represents continuous pixel values using quantum measurement expectations. Furthermore, this paper designs a quantum circuit architecture with structural inductive bias, where feature qubits controllably modulate a color qubit, allowing the color output to explicitly depend on the feature representation and thereby improving the model's capability in conditional pixel generation.

Our contributions are summarized as follows:
\begin{enumerate}
    \item We propose CoQui (\textbf{Co}ordinate-conditioned \textbf{Qu}antum \textbf{I}mplicit GAN), which formulates image generation as a continuous mapping from spatial coordinates to pixel intensities, rather than relying on the conventional amplitude-mapping paradigm. This design decouples the number of qubits from image resolution and mitigates the inter-pixel competition for probability mass inherent in conventional amplitude-based representations.

    \item We design a quantum circuit with structural inductive bias, in which feature qubits controllably modulate a color qubit. This mechanism makes the color generation process explicitly dependent on the input feature representation, thereby enhancing the modeling capability of the quantum circuit.

    \item We conduct systematic simulation experiments on multiple image generation benchmark datasets. Experimental results show that, compared with existing QGAN methods based on the amplitude-mapping paradigm, CoQui achieves better qualitative and quantitative performance and is able to generate images effectively with fewer qubits.
\end{enumerate}

\section{Related Work}\label{sec:related}

\subsection{Quantum Implicit Neural Representations and Image Generation}

In recent years, implicit neural representations (INRs) have attracted substantial attention because of their ability to represent high-quality signals by learning mappings from continuous coordinates to signal values \cite{Mescheder_2019_CVPR,sitzmann2020implicit,park2019deepsdf}. Unlike conventional discrete-grid representations \cite{gonzalez2009digital}, INRs model images \cite{chen2021learning,cao2023ciaosr}, audio \cite{sitzmann2020implicit,su2022inras}, and three-dimensional scenes \cite{zhao2024quantum, muller2022instant}  as continuous functions, from which signal values can be recovered through coordinate queries. Inspired by this idea, researchers have begun to investigate Quantum Implicit Neural Representations (QINRs) \cite{zhao2024quantum} built with parameterized quantum circuits (PQCs), aiming to exploit the potential of quantum circuits for high-dimensional nonlinear function approximation and Fourier-feature representation \cite{perez2020data,schuld2021effect}. 

Existing studies on QINRs mainly use them as continuous representers for conditional signals, with applications in deterministic mapping tasks such as image reconstruction \cite{zhao2024quantum,eren2026implementation,wang2026quantum}, compression \cite{fujihashi2025quantum}, and super-resolution \cite{jin2025qfgn,zhao2024quantum}. Although Zhang et al. proposed OQIDDM \cite{zhang2025denoising}, a QINR-based quantum diffusion model that introduces QINR into generative modeling, their method relies on amplitude encoding, which causes the required number of qubits to increase with image resolution. Thus, while QINRs show considerable promise for coordinate-dependent function representation, their integration with adversarial distribution learning remains insufficiently explored. To address this gap, this work investigates the feasibility of using QINR for end-to-end image distribution modeling within a Generative Adversarial Network (GAN) framework \cite{goodfellow2020generative}.

\subsection{Quantum Generative Adversarial Networks for Image Generation}

GANs learn real data distributions through a minimax game between a generator and a discriminator. To alleviate training instability and mode collapse, WGAN \cite{arjovsky2017wassersteingan} employs the Wasserstein distance to provide a smoother optimization objective, while WGAN-GP \cite{gulrajani2017improvedtrainingwassersteingans} further introduces a gradient penalty to enforce the Lipschitz constraint and improve training stability. This objective has also been adopted by many QGAN models to stabilize adversarial training. Quantum Generative Adversarial Networks (QGANs) extend adversarial generative modeling to the quantum computing framework. Early studies established the theoretical foundations of quantum adversarial learning \cite{lloyd2018quantum} and further used parameterized quantum circuits to implement trainable quantum generative models \cite{dallaire2018quantum}, with applications to numerical distribution modeling and quantum-state preparation \cite{zoufal2019quantum}.

More recently, QGANs have been applied to image generation. However, high-dimensional visual data impose substantial requirements on both the number of qubits and circuit depth. Existing methods therefore typically rely on dimensionality reduction or patch-based generation to alleviate scalability issues. Dimensionality-reduction approaches first compress images into a low-dimensional latent space using PCA or an encoder, then use a quantum generator to model the low-dimensional representation, and finally recover image dimensions through classical post-processing \cite{chu2023iqgan,silver2023mosaiq,vieloszynski2024latentqgan,chang2024latentstylebasedquantumgan,thomas2025vae}. Patch-based approaches divide an image into multiple local patches and use one or more quantum generators to synthesize the image blocks, thereby avoiding direct modeling of the full high-dimensional image \cite{huang2021experimental,tsang2023hybrid,yang2026ihqgan}. These methods reduce the training difficulty of quantum models, but they also weaken the direct role of the quantum generator in the complete image generation process. In particular, when classical decoders or multiple local generators are used, part of the generation capability may come from classical modules or task-specific structures.

To further improve the scalability of quantum image generation, recent studies have begun to explore end-to-end full-image quantum generators. Jäger et al. proposed a quantum Wasserstein GAN with a structural inductive bias \cite{jager2026scalingquantummachinelearning} . Their design follows the idea of an FRQI-like address-color image representation and generates full-resolution images without relying on dimensionality reduction or patch-based generation. Meanwhile, Yang et al. improved the amplitude decoding and prior-mapping mechanisms of the PQWGAN model, enabling end-to-end quantum image generation \cite{yang2026endtoendqganbasedimagesynthesis}.

We observe that most existing QGAN methods rely on quantum-state amplitudes to represent pixel intensities or feature values. This mechanism is constrained by the global normalization of quantum states and inevitably introduces numerical coupling among pixels, which weakens the model's ability to independently control local signal values. To address this limitation, this work explores an implicit quantum representation based on coordinate queries, where spatial coordinates and latent variables jointly drive a parameterized quantum circuit and pixel intensities are directly extracted from measurement expectations of observables. This formulation provides a more flexible route toward end-to-end image distribution modeling.

\section{Method}

\subsection{Problem Formulation}

This work proposes CoQui, a QGAN framework based on Quantum Implicit Neural Representation (QINR). Instead of generating an image by assigning pixel intensities to quantum-state amplitudes, CoQui reformulates image generation as a continuous implicit mapping conditioned jointly on spatial coordinates and a latent variable.

CoQui represents a grayscale image as
$X\in[0,1]^{H\times W\times 1}$ and learns a coordinate-conditioned
generator $G_{\Phi}(c,z)\in[0,1]$. Here,
$c=(x,y)\in\mathcal{C}\subset[0,1)^2$ is a normalized pixel coordinate,
$z\in\mathbb{R}^{d_z}$ is a latent code shared across all coordinates of
one image, and $\Phi$ denotes the trainable parameters. A complete image
is obtained by evaluating the generator over
$\mathcal{C}_{H,W}=\{(i/W,j/H)\mid 0\leq i<W,\ 0\leq j<H\}$.

\subsection{Conditional Input to the Quantum Implicit Generator}

To improve the ability of coordinate inputs to represent local details and high-frequency structures, the two-dimensional coordinate $c=(x,y)$ is first transformed by positional encoding. Given the number of frequencies $K$, the encoding function is defined as
\begin{equation}
\begin{aligned}
\gamma(c)
&=\Bigl[
x,y,\{
\sin(2^k x),\cos(2^k x),\\
&\qquad\qquad
\sin(2^k y),\cos(2^k y)
\}_{k=0}^{K-1}
\Bigr].
\end{aligned}
\end{equation}
with encoded coordinate dimension $d_{\gamma}=2+4K$.
The encoded coordinate feature is concatenated with the latent variable $z\in\mathbb{R}^{d_z}$:
\begin{equation}
u(c,z)=[\gamma(c),z]\in\mathbb{R}^{d_{\gamma}+d_z}.
\end{equation}
This vector is fed into a classical embedding network $\Gamma$ composed of fully connected layers and ReLU activations. The network outputs $3N_f$ values, where $N_f$ is the number of feature qubits. A hyperbolic tangent nonlinearity and an angle scaling factor $\alpha$ are then applied to obtain the base rotation angles:
\begin{equation}
a(c,z)=\alpha\tanh(\Gamma(u(c,z)))\in\mathbb{R}^{3N_f}.
\end{equation}
The vector is reshaped as
\begin{equation}
A(c,z)=\operatorname{Reshape}(a(c,z))\in\mathbb{R}^{N_f\times 3}.
\end{equation}
For the $i$-th feature qubit, the three base angles are
\begin{equation}
\begin{aligned}
A_i(c,z)
&=\left(
a_{i,x}(c,z),a_{i,y}(c,z),a_{i,z}(c,z)
\right),\\
&\qquad i=1,\ldots,N_f.
\end{aligned}
\end{equation}
which are used for $R_X$, $R_Y$, and $R_Z$ input rotations each.

\subsection{Quantum Implicit Generator Circuit}

The quantum generator consists of $N_q=N_f+1$ qubits. The first qubit $q_0$ is designated as the color qubit for pixel readout, while the remaining qubits $q_1,\ldots,q_{N_f}$ are feature qubits that encode coordinate- and latent-dependent implicit features. The circuit is initialized from the all-zero state.

\subsubsection{Color-Qubit Brightness Initialization}
Initially, a trainable $R_Y$ rotation is applied to the color qubit:
\begin{equation}
U_{\mathrm{bias}}=R_Y^{(q_0)}(\beta),
\end{equation}
where $\beta$ is a trainable brightness-bias parameter. To give the model a reasonable initial pixel-brightness distribution, $\beta$ is initialized according to a preset mean pixel value $\mu_0\in[0,1]$:
\begin{equation}
\beta_0=\arccos(1-2\mu_0).
\end{equation}
Ignoring subsequent feature-to-color writing operations, the Pauli-$Z$ expectation of the color qubit is $\langle Z_0\rangle=\cos\beta_0$, and the corresponding initial pixel value is
\begin{equation}
\frac{1-\langle Z_0\rangle}{2}=\frac{1-\cos\beta_0}{2}=\mu_0.
\end{equation}
This initialization explicitly controls the initial average brightness of the generator and improves early-stage training stability.

\subsubsection{Scaled Data Re-uploading Layer}
The quantum circuit comprises $L$ data re-uploading layers. Rather than injecting the same base angles $A(c,z)$ identically at every layer, CoQui equips each layer $l$ with trainable scale and bias parameters:
$S_{l}\in\mathbb{R}^{N_f\times 3}, B_{l}\in\mathbb{R}^{N_f\times 3}$. The angles injected at layer $l$ are obtained through the element-wise affine transformation
\begin{equation}
\Theta_{l}(c,z)=S_{l}\odot A(c,z)+B_{l},
\end{equation}
where $\odot$ denotes element-wise multiplication. Specifically, for feature qubit $q_i$ and rotation axis $r\in\{x,y,z\}$,
\begin{equation}
\theta_{i,r}^{l}(c,z)
=s_{i,r}^{l}a_{i,r}(c,z)+b_{i,r}^{l}.
\end{equation}
Here, $s_{i,r}^{l}$ controls the strength of the coordinate--latent signal, whereas $b_{i,r}^{l}$ provides an input-independent angular offset. The three angles associated with $q_i$ are grouped as
\begin{equation}
\Theta_{i,l}(c,z)
=\left(
\theta_{i,1}^{l}(c,z),
\theta_{i,2}^{l}(c,z),
\theta_{i,3}^{l}(c,z)
\right).
\end{equation}
This encoding acts on all feature qubits:
\begin{equation}
\begin{aligned}
U_{\mathrm{enc}}^{l}
&=\prod_{i=1}^{N_f}
R_Z^{(q_i)}\!\left(\theta_{i,z}^{l}(c,z)\right)
R_Y^{(q_i)}\!\left(\theta_{i,y}^{l}(c,z)\right)\\
&\qquad\quad
R_X^{(q_i)}\!\left(\theta_{i,x}^{l}(c,z)\right).
\end{aligned}
\end{equation}

This layer-dependent reparametrization lets each layer emphasize, attenuate, or shift components of the shared coordinate--latent representation, enhancing the PQC's expressive capacity for implicit function modeling.

\subsubsection{Feature-Qubit Local Rotations}
After input encoding, each feature qubit receives trainable local rotations with parameters $W_{l}\in\mathbb{R}^{N_f\times 3}$ %
where $W_i^{l}=(w_{i,1}^{l},w_{i,2}^{l},w_{i,3}^{l})$ for the $i$-th feature qubit. The local rotation unit is
\begin{equation}
U_{\mathrm{loc}}^{l}=\prod_{i=1}^{N_f} R_Y^{(q_i)}(w_{i,3}^{l})R_Z^{(q_i)}(w_{i,2}^{l})R_Y^{(q_i)}(w_{i,1}^{l}).
\end{equation}
This component provides input-independent trainable quantum transformations after coordinate-latent encoding.

\subsubsection{Feature-Feature Entanglement}
To model correlations among feature qubits, CoQui introduces a ring CNOT entanglement pattern:
\begin{equation}
U_{\mathrm{ent}}^{l}=\prod_{i=1}^{N_f}\operatorname{CNOT}(q_i,q_{i+1}),\qquad q_{N_f+1}\equiv q_1.
\end{equation}

\subsubsection{Feature-to-Color Writing}
The core circuit design writes information from feature qubits to the color qubit through controlled-$R_Y$ gates. For layer $l$ and feature qubit $q_i$, let $\eta_i^{l}\in\mathbb{R}$ denote the trainable color-writing parameter. The feature-to-color operation is $\operatorname{CRY}_{q_i\rightarrow q_0}(\eta_i^{l})$, where $q_i$ is the control qubit and $q_0$ is the target qubit. The complete writing module at layer $l$ is
\begin{equation}
U_{\mathrm{write}}^{l}=\prod_{i=1}^{N_f}\operatorname{CRY}_{q_i\rightarrow q_0}(\eta_i^{l}).
\end{equation}
This module makes the color-qubit state depend explicitly on feature qubits that encode coordinate and latent information. Thus, feature qubits form implicit spatial features, while the color qubit acts as the measurable pixel-output channel.

\subsubsection{Color Residual Update}
After each feature-to-color writing module, CoQui applies a trainable residual update on the color qubit:
\begin{equation}
U_{\mathrm{res}}^{l}=R_Y^{(q_0)}(r_{l,3})R_Z^{(q_0)}(r_{l,2})R_Y^{(q_0)}(r_{l,1}),
\end{equation}
where $r_{l}=(r_{l,1},r_{l,2},r_{l,3})$ denotes the color residual parameters at layer $l$. This module provides additional local degrees of freedom for the readout qubit, allowing pixel intensities to be accumulated, adjusted, and refined across multiple re-uploading layers.

Combining the above modules, the $l$-th quantum transformation is
\begin{equation}
U_{l}(c,z)=U_{\mathrm{res}}^{l}U_{\mathrm{write}}^{l}U_{\mathrm{ent}}^{l}U_{\mathrm{loc}}^{l}U_{\mathrm{enc}}^{l}(c,z).
\end{equation}
For coordinate $c=(x,y)$ and latent variable $z$, the complete PQC outputs
\begin{equation}
|\psi_{\Theta}(c,z)\rangle=U_L(c,z)\cdots U_2(c,z)U_1(c,z)U_{\mathrm{bias}}|\psi_0\rangle.
\end{equation}
The generator parameters are $\Phi=\{s^{l},b^{l},W^{l},\eta^{l},r^{l}\}_{l=1}^{L}$.

\subsection{Quantum Measurement and Pixel Readout}

After $L$ layers of quantum evolution, CoQui measures only the Pauli-$Z$ expectation of the color qubit:
\begin{equation}
m_{\Phi}(c,z)=\langle Z_0\rangle=\langle\psi_{\Phi}(c,z)|Z_0|\psi_{\Phi}(c,z)\rangle.
\end{equation}
The expectation is then mapped linearly to a normalized grayscale pixel intensity:
\begin{equation}
G_{\Phi}(c,z)=\frac{1-m_{\Phi}(c,z)}{2}.
\end{equation}
Since $m_{\Phi}(c,z)\in[-1,1]$, the output satisfies $G_{\Phi}(c,z)\in[0,1]$. This design uses a physical observable of the color qubit directly as the pixel output, making the quantum circuit itself responsible for pixel-intensity modeling.

Given a latent variable $z$, a full image is generated by querying all coordinates in $\mathcal{C}_{H,W}$:
\begin{equation}
\begin{aligned}
\hat{X}_{j,i}
&=G_{\Phi}\left(\frac{i}{W},\frac{j}{H},z\right),
 0\le i<W, 0\le j<H .
\end{aligned}
\end{equation}
yielding
\begin{equation}
\hat{X}=G_{\Phi}(z)\in[0,1]^{H\times W\times 1}.
\end{equation}

\subsection{Training Objective}

Following PQWGAN \cite{tsang2023hybrid}, CoQui is trained using the Wasserstein GAN objective with gradient penalty. The critic distinguishes real images from generated ones, while the generator learns to produce realistic images. Further implementation details of the classical embedding network and the critic are provided in \ref{app:classical_components}.

\section{Experiments}

\subsection{Experimental Setup}

We evaluate CoQui on MNIST and Fashion-MNIST at the standard $28\times 28$ resolution. The training dataset consists of 1000 real images in each experiment. The model is trained with Adam for 1000 epochs with a batch size of 5. Following the setting of Jäger et al., the learning rates of the generator and discriminator are set to $1\times10^{-3}$ and $1\times10^{-4}$, respectively.
 The quantum generator used in the experiments contains one color qubit and $N_f=4$ feature qubits, and the circuit depth is set to $r=20$ re-uploading layers.
The hyperparameters of the other classical components are reported in \ref{app:classical_components}.
All experiments are performed in simulation on a machine with an AMD Ryzen 9 9950X3D 16-Core Processor at 4.30 GHz and 64 GB RAM. All models are implemented in JAX.

\paragraph{Baselines.}

We compare CoQui with two representative quantum image generation methods, PQWGAN and the end-to-end Quantum Wasserstein QGAN proposed by Jäger et al., both of which adopt the amplitude-mapping paradigm. Experiments are conducted under two evaluation settings. We first compare the methods on the standard multi-class benchmark following the protocol of Jäger et al. We then perform per-class experiments, where each method is trained independently on each class using the same training protocol and data budget. This complementary evaluation assesses both the ability to model a heterogeneous image distribution and the generation quality under matched training conditions. In addition, we implement a classical counterpart of CoQui to isolate the contribution of the proposed quantum generator. All baseline methods are reimplemented by us and optimized for efficient classical simulation. The code will be made publicly available upon acceptance.

\begin{figure}[t]
\centering
\begin{minipage}{\linewidth}
\centering
\makebox[\linewidth][c]{%
\includegraphics[width=0.98\linewidth]{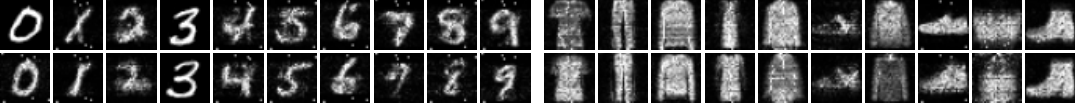}}
\par\vspace{-0.6mm}
{\scriptsize PQWGAN\par}
\end{minipage}\par
\begin{minipage}{\linewidth}
\centering
\makebox[\linewidth][c]{%
\includegraphics[width=0.98\linewidth]{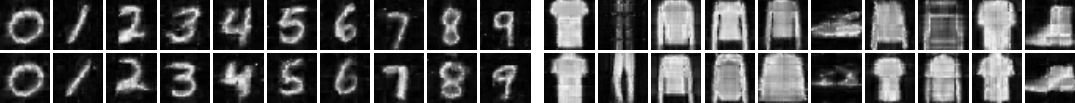}}
\par\vspace{-0.6mm}
{\scriptsize Wasserstein QGAN \par}
\end{minipage}\par
\begin{minipage}{\linewidth}
\centering
\makebox[\linewidth][c]{%
\includegraphics[width=0.98\linewidth]{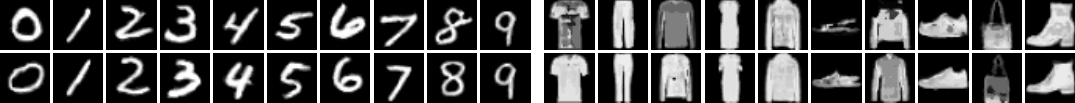}}
\par\vspace{-0.6mm}
{\scriptsize Classical INR-GAN\par}
\end{minipage}\par
\begin{minipage}{\linewidth}
\centering
\makebox[\linewidth][c]{%
\includegraphics[width=0.98\linewidth]{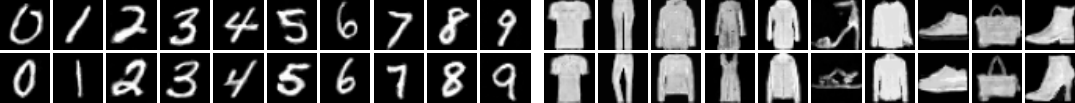}}
\par\vspace{-0.6mm}
{\scriptsize CoQui\par}
\end{minipage}\par
\vspace{-1mm}
\begingroup
\setlength{\tabcolsep}{0pt}
\renewcommand{\arraystretch}{1.08}
\newcommand{\samplegrid}[2]{%
  \includegraphics[
    width=0.19\columnwidth,
    height=0.2375\columnwidth,
    trim={#1},
    clip
  ]{#2}%
}
\newcommand{\samplegridwithgaps}[2]{%
  \begin{tikzpicture}[
    x=0.19\columnwidth,
    y=0.2375\columnwidth,
    baseline=0pt
  ]
    \node[anchor=south west,inner sep=0] at (0,0) {%
      \includegraphics[
        width=0.19\columnwidth,
        height=0.2375\columnwidth,
        trim={#1},
        clip
      ]{#2}%
    };
    \foreach \x in {0.25,0.5,0.75}
      \draw[white,line width=0.6pt] (\x,0) -- (\x,1);
    \foreach \y in {0.2,0.4,0.6,0.8}
      \draw[white,line width=0.6pt] (0,\y) -- (1,\y);
  \end{tikzpicture}%
}
\newcommand{\methodlabel}[1]{%
  \parbox[c][14.5pt][c]{0.19\columnwidth}{%
    \centering\fontsize{5.5}{6}\selectfont #1\par
  }%
}
\newcommand{\fidlabel}[1]{%
  \makebox[0.19\columnwidth][c]{%
    \fontsize{5.5}{6}\selectfont FID \textbf{#1}%
  }%
}
\begin{tabular}{@{}c@{\hspace{2pt}}c@{\hspace{2pt}}c@{\hspace{2pt}}c@{\hspace{2pt}}c@{}}
\fidlabel{221} &
\fidlabel{118} &
\fidlabel{195} &
\fidlabel{34} &
\fidlabel{42}
\\[-1mm]
\samplegrid{7pt 7pt 7pt 7pt}{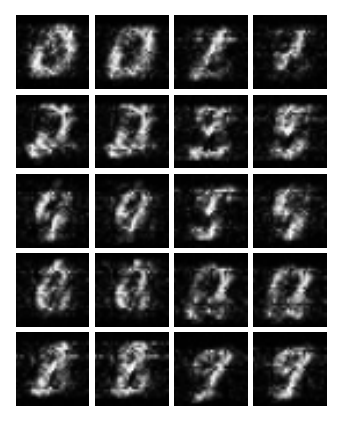} &
\samplegrid{7pt 7pt 7pt 7pt}{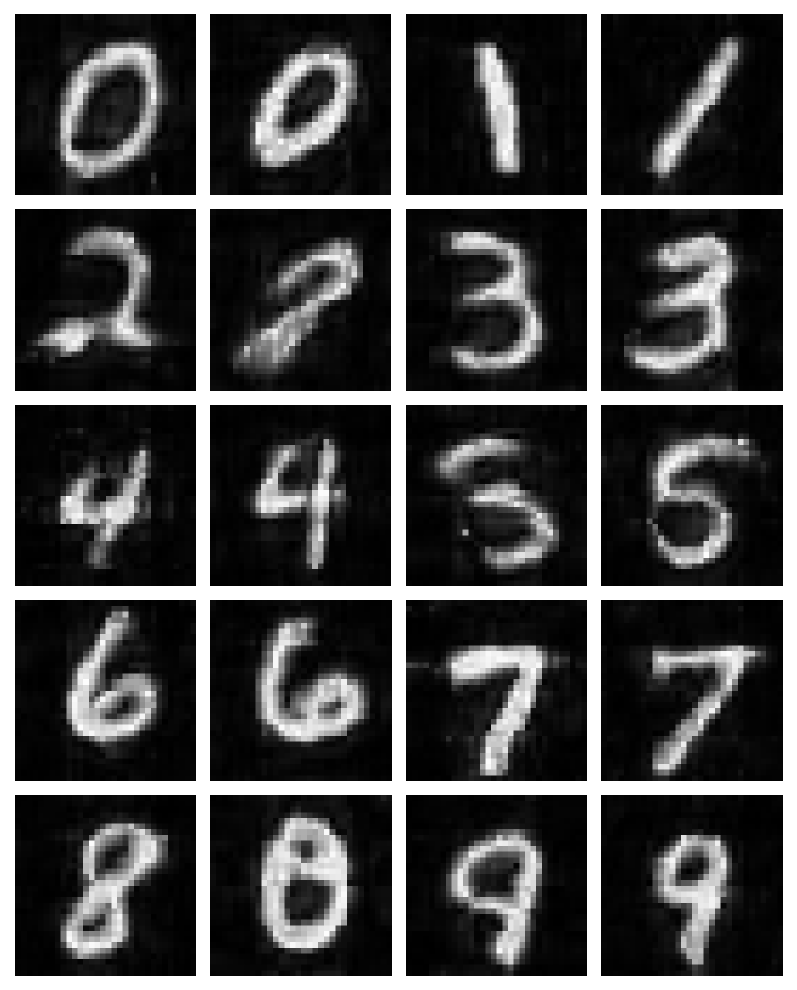} &
\samplegrid{10pt 5pt 10pt 6pt}{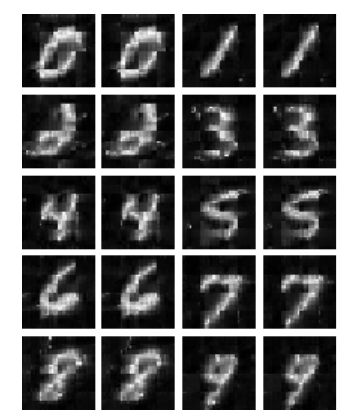} &
\samplegrid{7pt 5pt 7pt 6pt}{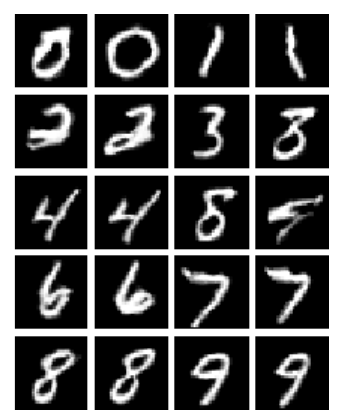} &
\samplegrid{7pt 7pt 7pt 7pt}{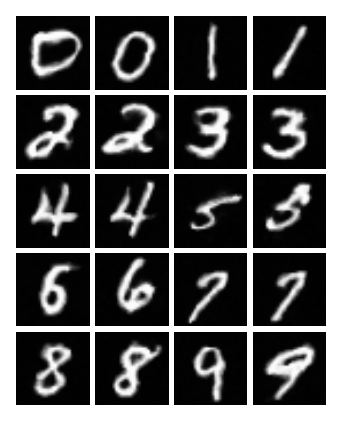}
\\[-2mm]
\fidlabel{244} &
\fidlabel{91} &
\fidlabel{176} &
\fidlabel{72} &
\fidlabel{72}
\\[-1mm]
\samplegrid{7pt 7pt 7pt 7pt}{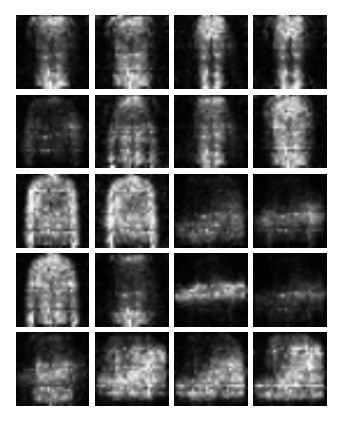} &
\samplegridwithgaps{7pt 7pt 7pt 7pt}{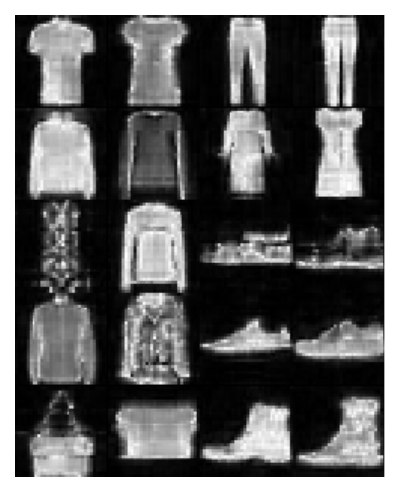} &
\samplegrid{7pt 6pt 7pt 6pt}{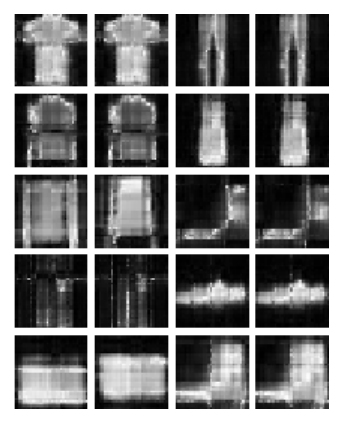} &
\samplegrid{7pt 5pt 7pt 6pt}{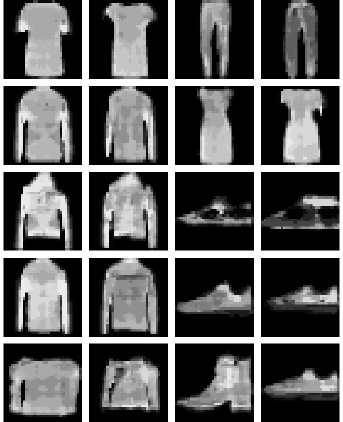} &
\samplegrid{7pt 7pt 7pt 7pt}{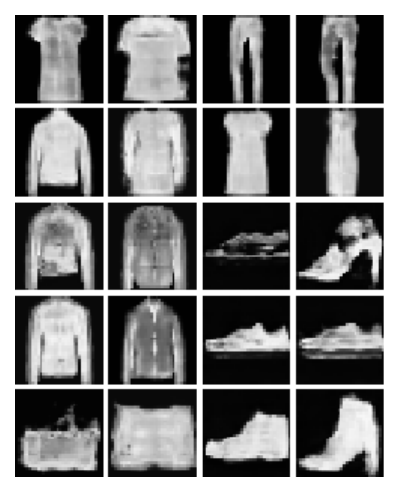}
\\[-1mm]
\methodlabel{PQWGAN} &
\methodlabel{Wasserstein\\QGAN\\(Original setting)} &
\methodlabel{Wasserstein\\QGAN\\} &
\methodlabel{Classical\\INR-GAN} &
\methodlabel{CoQui}
\end{tabular}
\endgroup
\caption{Qualitative comparison on MNIST and Fashion-MNIST. Top: samples from each class for each method. Bottom: representative grids over ten classes, with FID scores reported above each grid.}
\label{fig:qualitative_comparison}
\end{figure}
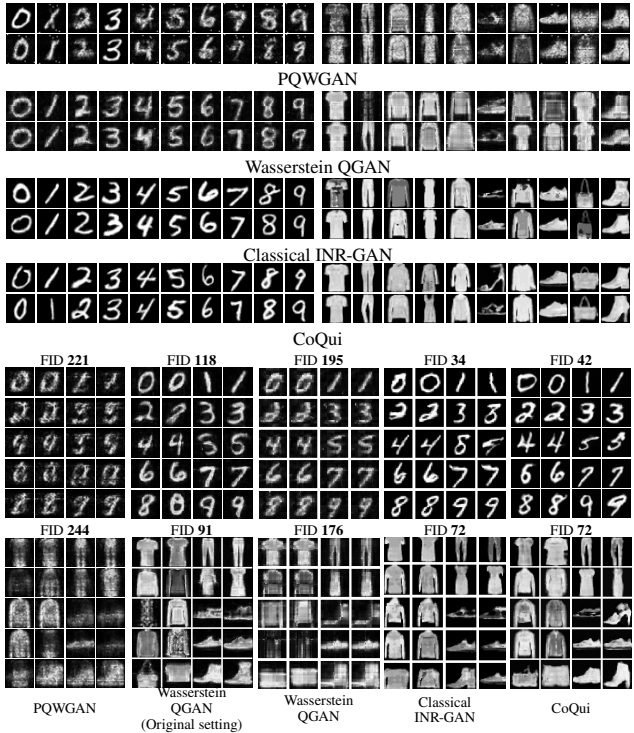
\paragraph{Evaluation Metric.}
We evaluate generated samples using four complementary metrics: FID, classifier-based diversity metrics, feature-space precision and recall, and auxiliary low-level visual statistics. \textbf{FID} measures the overall discrepancy between real and generated distributions~\cite{jager2026scalingquantummachinelearning}. To evaluate class diversity, a pretrained MNIST/Fashion-MNIST classifier is used to obtain the predicted class distribution $p_g(y)$, following the classifier-based evaluation idea of Inception Score and Mode Score~\cite{salimans2016improved,che2016mode}. We report \textbf{Entropy} as $H_{\mathrm{norm}}(p_g)=-\frac{1}{\log C}\sum_{c=1}^{C}p_g(c)\log p_g(c)$, where $C=10$, and \textbf{JSD} between $p_g(y)$ and the uniform distribution $u(y)$, defined as $\operatorname{JSD}(p_g\|u)=\frac{1}{2}\operatorname{KL}(p_g\|m)+\frac{1}{2}\operatorname{KL}(u\|m)$ with $m=\frac{1}{2}(p_g+u)$~\cite{lin1991divergence}. We further use \textbf{P@5} and \textbf{R@5} to measure sample fidelity and distribution coverage in the Inception feature space, following the improved precision and recall metric for generative models~\cite{kynkaanniemi2019improved}. Finally, \textbf{Brightness} and \textbf{Sobel} are defined as $D_{\mathrm{bright}}=|\operatorname{mean}(x_g)-\operatorname{mean}(x_r)|$ and $D_{\mathrm{sobel}}=|\operatorname{mean}(\operatorname{Sobel}(x_g))-\operatorname{mean}(\operatorname{Sobel}(x_r))|$, respectively, where the Sobel operator is used to estimate image gradient magnitude~\cite{sobel2014isotropic}. Lower values are better for FID, JSD, Brightness, and Sobel, whereas higher values are better for Entropy, P@5, and R@5.

\setcounter{table}{0}
\begin{table}[H]
\centering
\scriptsize
\begingroup
\renewcommand{\arraystretch}{1}
\setlength{\tabcolsep}{1.2pt}
\resizebox{\columnwidth}{!}{%
\begin{tabular}{@{}cc*{10}{c}@{}}
\toprule
\rowcolor{headergray}
Dataset & Method & 0 & 1 & 2 & 3 & 4 & 5 & 6 & 7 & 8 & 9 \\
\midrule
\multirow{4}{*}{MNIST}
& PQWGAN & 210 & 137 & 223 & 203 & 191 & 187 & 179 & 179 & 195 & 199 \\
& \makecell[c]{Wasserstein QGAN} & 214 & 122 & 203 & 174 & 189 & 175 & 184 & 172 & 193 & 197 \\
& \makecell[c]{Classical\\INR-GAN} & 20 & \textbf{16} & \textbf{22} & \textbf{14} & 21 & \textbf{19} & \textbf{16} & \textbf{20} & \textbf{19} & 19 \\
& CoQui & \textbf{17} & 17 & 26 & 17 & \textbf{19} & 21 & 17 & 25 & \textbf{19} & \textbf{16} \\
\midrule
\multirow{4}{*}{\makecell[c]{Fashion-\\MNIST}}
& PQWGAN & 241 & 97 & 204 & 175 & 213 & 210 & 192 & 219 & 275 & 230 \\
& \makecell[c]{Wasserstein QGAN} & 167 & 112 & 164 & 183 & 162 & 169 & 165 & 246 & 184 & 176 \\
& \makecell[c]{Classical\\INR-GAN} & 53 & \textbf{25} & 54 & 44 & 56 & \textbf{58} & 61 & \textbf{43} & \textbf{47} & 39 \\
& CoQui & \textbf{47} & 28 & \textbf{45} & \textbf{40} & \textbf{41} & 62 & \textbf{52} & \textbf{43} & 55 & \textbf{38} \\
\bottomrule
\end{tabular}%
}
\endgroup
\caption{Comparison on MNIST and Fashion-MNIST. Best results are in bold.}
\label{tab:mnist_fid_comparison}
\end{table}
\setcounter{table}{1}
\begin{table}[htbp]
\centering
\scriptsize
\setlength{\tabcolsep}{4pt}
\begin{tabular}{lcc}
\toprule
\rowcolor{headergray}
Method & Qubits & Number of Quantum Circuits  \\
\midrule
PQWGAN & $192$ & $32$   \\
Wasserstein QGAN & $11$ & 1  \\
CoQui & $5$ & 1  \\
\bottomrule
\end{tabular}
\caption{Comparison of quantum resource requirements across different QGAN models.}
\label{tab:baseline}
\end{table}
\subsection{Main Results}
Figure~\ref{fig:qualitative_comparison} shows representative generated samples on class-wise subsets and on the full dataset containing ten classes.
The FID scores for the full-dataset setting are directly annotated in the figure, while the class-wise FID scores are reported in Table~\ref{tab:mnist_fid_comparison}. On the class-wise subsets, CoQui generates samples with clearly recognizable category structure and good visual quality. This is especially evident on the more complex Fashion-MNIST dataset, where CoQui still produces visually meaningful samples. In contrast, the two amplitude-mapping based QGAN models generate noticeably blurrier samples. This is mainly due to the pixel coupling introduced by amplitude encoding: pixels share normalized amplitudes and are therefore difficult to adjust independently, which limits the representation of fine-grained features.
Compared with the Classical INR-GAN of a similar parameter scale, CoQui achieves the best or tied-best FID scores on four MNIST classes and seven Fashion-MNIST classes. This indicates that CoQui retains strong generative capability under a low quantum-resource budget, highlighting the potential of quantum generative models for image generation. The full-dataset setting with ten classes is more challenging than class-wise generation because it involves a more complex data distribution; accordingly, all methods show degraded performance.
Nevertheless, PQWGAN gives the weakest visual results, followed by Wasserstein QGAN trained under the same setting, while Wasserstein QGAN trained with the original 6K-sample setting shows some improvement.
During our reproduction of Wasserstein QGAN, we observed that the original work employs a post-selection strategy for visualization: for each noise pattern, 500 candidate samples are generated, and the sample with the smallest Euclidean distance to their mean is selected for display. We do not apply this selection procedure and instead directly visualize randomly generated samples.
Compared with Classical INR-GAN, CoQui achieves comparable overall visual quality and shows advantages in local detail modeling.

In addition, as shown in Table~\ref{tab:baseline}, CoQui uses only five qubits and one quantum circuit, giving it the lowest quantum-resource cost among all compared methods. The qubit requirement of amplitude-mapping based methods typically grows with image resolution, whereas the qubit count of CoQui is not directly tied to image resolution. CoQui therefore provides better resource efficiency and scalability. The training convergence behavior is analyzed in ~\ref{app:Training Convergence Analysis}.

\subsection{Ablation Study 1: circuit architecture ablation}
We further conduct ablation studies to examine the contribution of the proposed circuit architecture and the color-qubit design. Figure~\ref{fig:mnist_ablation_samples} provides representative MNIST samples for these variants, while Tables~\ref{tab:hardware_efficient_comparison} and~\ref{tab:ablation_results} report the corresponding quantitative results.

\subsubsection{Circuit design ablation}
Table~\ref{tab:hardware_efficient_comparison} compares CoQui using the proposed feature-to-color circuit with a hardware-efficient ansatz. The proposed circuit improves most metrics, reducing FID from $41.41$ to $40.15$ and JSD from $0.00858$ to $0.00411$, while increasing entropy from $0.9856$ to $0.9930$ and P@5 from $0.489$ to $0.505$. It also yields lower brightness and Sobel discrepancies, indicating better low-level image statistics and sharper local structures. The hardware-efficient circuit gives a slightly higher R@5, but its weaker FID, JSD, and edge statistics suggest poorer overall distributional alignment. These results show that the structured feature-to-color modulation is important for stable and faithful image generation.

\setcounter{table}{2}
\begin{table}[t]
\centering
\scriptsize
\renewcommand{\arraystretch}{0.95}
\setlength{\tabcolsep}{2pt}
\begin{tabular}{
@{}
p{0.16\columnwidth}
>{\centering\arraybackslash}p{0.33\columnwidth}
>{\centering\arraybackslash}p{0.33\columnwidth}
@{}
}
\toprule
\rowcolor{headergray}
\textbf{Metric}
&
\makecell[c]{CoQui with\\
hardware-efficient circuit}
&
\makecell[c]{CoQui with\\
our proposed circuit}
\\
\midrule
FID $\downarrow$ & $41.41 \pm 6.10$ & $\mathbf{40.15 \pm 3.45}$ \\
JSD $\downarrow$ & $0.00858 \pm 0.00333$ & $\mathbf{0.00411 \pm 0.00128}$ \\
Entropy $\uparrow$ & $0.9856 \pm 0.0055$ & $\mathbf{0.9930 \pm 0.0021}$ \\
P@5 $\uparrow$ & $0.489 \pm 0.081$ & $\mathbf{0.505 \pm 0.061}$ \\
R@5 $\uparrow$ & $\mathbf{0.485 \pm 0.250}$ & $0.467 \pm 0.108$ \\
Brightness $\downarrow$ & $0.0065 \pm 0.0046$ & $\mathbf{0.0058 \pm 0.0051}$ \\
Sobel $\downarrow$ & $0.0200 \pm 0.0000$ & $\mathbf{0.0073 \pm 0.0033}$ \\
\bottomrule
\end{tabular}
\caption{Quantitative comparison between CoQui with the hardware-efficient circuit
and CoQui with our proposed circuit. Best results are shown in bold.}
\label{tab:hardware_efficient_comparison}
\end{table}

\subsubsection{Core-component ablation}
Figure~\ref{fig:mnist_ablation_samples}(b) and Table~\ref{tab:ablation_results} evaluate the main color-qubit components. Full CoQui achieves the best JSD, entropy, brightness discrepancy, and Sobel discrepancy, showing the most balanced match to the real distribution. Removing the color-qubit residual or the color-bias initialization can slightly improve FID, but both variants worsen low-level statistics, especially edge consistency. For example, removing the residual increases the Sobel discrepancy from $0.0073$ to $0.0370$. Removing the scaled data re-uploading causes the largest degradation, increasing FID from $40.15$ to $56.54$, which indicates that layer-wise input scaling is important for effectively modulating coordinate- and noise-dependent quantum features. Overall, these ablations suggest that the complete color-qubit design is not optimized for a single metric only, but provides the best trade-off across distributional fidelity, sample diversity, brightness, and local structure.
\begingroup
\setlength{\intextsep}{2pt}
\setlength{\textfloatsep}{2pt}
\setlength{\floatsep}{2pt}
\begin{figure}[H]
    \centering
    \begin{minipage}{0.60\columnwidth}
        \centering
        \includegraphics[
            width=\linewidth,
            trim=0 58.76pt 0 0,
            clip
        ]{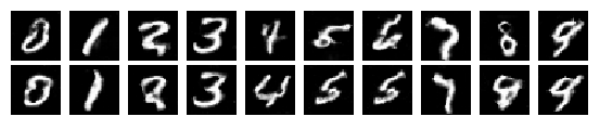}
        \par\vspace{-1.5mm}
        {\scriptsize CoQui with hardware-efficient ansatz\par}
        \vspace{0.6mm}
        \includegraphics[
            width=\linewidth,
            trim=0 0 0 58.76pt,
            clip
        ]{mnist_baseline_vs_hardware_image_only_representative.pdf}
        \par\vspace{-1.5mm}
        {\scriptsize CoQui with our proposed circuit\par}
        \par\vspace{-0.1mm}
        {\scriptsize\textbf{(a)} Circuit ablation\par}
    \end{minipage}
    \par\vspace{0.1mm}
    \begin{minipage}{\columnwidth}
        \centering
        \begin{tabular}{@{}c@{\hspace{0.4mm}}c@{}}
            \begin{minipage}[t]{0.49\linewidth}
                \centering
                \includegraphics[
                    width=\linewidth,
                    trim=0 152.88pt 0 0,
                    clip
                ]{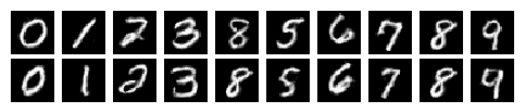}
                \par\vspace{-1.5mm}
                {\scriptsize Full CoQui \par}
            \end{minipage}
            &
            \begin{minipage}[t]{0.49\linewidth}
                \centering
                \includegraphics[
                    width=\linewidth,
                    trim=0 101.92pt 0 50.96pt,
                    clip
                ]{mnist_core_ablation_image_only_representative.pdf}
                \par\vspace{-1.5mm}
                {\scriptsize w/o color-qubit residual\par}
            \end{minipage}
            \\[-0.2mm]
            \begin{minipage}[t]{0.49\linewidth}
                \centering
                \includegraphics[
                    width=\linewidth,
                    trim=0 50.96pt 0 101.92pt,
                    clip
                ]{mnist_core_ablation_image_only_representative.pdf}
                \par\vspace{-1.5mm}
                {\scriptsize w/o color-qubit bias initialization\par}
            \end{minipage}
            &
            \begin{minipage}[t]{0.49\linewidth}
                \centering
                \includegraphics[
                    width=\linewidth,
                    trim=0 0 0 152.88pt,
                    clip
                ]{mnist_core_ablation_image_only_representative.pdf}
                \par\vspace{-1.5mm}
                {\scriptsize w/o scaled data re-uploading\par}
            \end{minipage}
        \end{tabular}
        \par\vspace{-0.1mm}
        {\scriptsize\textbf{(b)} Core-component ablation\par}
    \end{minipage}

    \caption{Representative MNIST samples for (a) CoQui and its
    hardware-efficient variant, and (b) the CoQui baseline and
    variants without the color residual, color-bias initialization,
    or scaled data re-uploading.}
    \label{fig:mnist_ablation_samples}
    \label{fig:mnist_baseline_vs_hardware}
    \label{fig:mnist_core_ablation}
\end{figure}

\setcounter{table}{3}
\begin{table}[H]
\centering
\scriptsize
\renewcommand{\arraystretch}{0.95}
\setlength{\tabcolsep}{0pt}
\begin{tabular}{
@{}
p{0.16\columnwidth}
>{\centering\arraybackslash}p{0.18\columnwidth}
>{\centering\arraybackslash}p{0.18\columnwidth}
>{\centering\arraybackslash}p{0.18\columnwidth}
>{\centering\arraybackslash}p{0.18\columnwidth}
@{}
}
\toprule
\rowcolor{headergray}
Metric
& \makecell[c]{Full\\CoQui}
& \makecell[c]{CoQui w/o\\color-qubit\\residual}
& \makecell[c]{CoQui w/o\\color-bias\\initialization}
& \makecell[c]{CoQui w/o\\scaled data\\ re-uploading } \\
\midrule
FID $\downarrow$
& \makecell[c]{$40.15$\\[-0.35mm]$\pm 3.45$}
& \makecell[c]{$38.82$\\[-0.35mm]$\pm 1.52$}
& \makecell[c]{$\mathbf{36.81}$\\[-0.35mm]$\mathbf{\pm 3.30}$}
& \makecell[c]{$56.54$\\[-0.35mm]$\pm 3.63$} \\
JSD $\downarrow$
& \makecell[c]{$\mathbf{0.00411}$\\[-0.35mm]$\mathbf{\pm 0.00128}$}
& \makecell[c]{$0.00493$\\[-0.35mm]$\pm 0.00502$}
& \makecell[c]{$0.00560$\\[-0.35mm]$\pm 0.00079$}
& \makecell[c]{$0.00439$\\[-0.35mm]$\pm 0.00063$} \\
Entropy $\uparrow$
& \makecell[c]{$\mathbf{0.9930}$\\[-0.35mm]$\mathbf{\pm 0.0021}$}
& \makecell[c]{$0.9915$\\[-0.35mm]$\pm 0.0087$}
& \makecell[c]{$0.9903$\\[-0.35mm]$\pm 0.0015$}
& \makecell[c]{$0.9924$\\[-0.35mm]$\pm 0.0009$} \\
Brightness $\downarrow$
& \makecell[c]{$\mathbf{0.0058}$\\[-0.35mm]$\mathbf{\pm 0.0051}$}
& \makecell[c]{$0.0133$\\[-0.35mm]$\pm 0.0015$}
& \makecell[c]{$0.0084$\\[-0.35mm]$\pm 0.0045$}
& \makecell[c]{$0.0091$\\[-0.35mm]$\pm 0.0033$} \\
Sobel $\downarrow$
& \makecell[c]{$\mathbf{0.0073}$\\[-0.35mm]$\mathbf{\pm 0.0033}$}
& \makecell[c]{$0.0370$\\[-0.35mm]$\pm 0.0200$}
& \makecell[c]{$0.0142$\\[-0.35mm]$\pm 0.0171$}
& \makecell[c]{$0.0182$\\[-0.35mm]$\pm 0.0036$} \\
\bottomrule
\end{tabular}

\caption{Quantitative core-component ablation results for CoQui. Best results are shown in bold.}
\label{tab:ablation_results}
\end{table}

\subsection{Ablation Study 2: Model Capacity Analysis}

Figure~\ref{fig:mnist_nf_ablation} and Table~\ref{tab:mnist_qubit_selection} evaluate the effect of model capacity from two aspects: the number of feature qubits and the circuit depth. Since the computational cost of quantum circuit simulation grows rapidly with both the number of qubits and the circuit depth, we first identify a suitable number of feature qubits under a fixed circuit depth, and then compare different circuit depths based on this setting.
\setcounter{table}{4}
\begin{table}[H]\label{tab:capacity}
\centering
\scriptsize
\renewcommand{\arraystretch}{0.95}
\setlength{\tabcolsep}{5pt}
\begin{tabular}{@{}llccc@{}}
\toprule
\rowcolor{headergray}
Ablation Study & Setting & FID $\downarrow$ & Class JSD $\downarrow$ & Class Entropy $\uparrow$ \\
\midrule
\multirow{5}{*}{\makecell[c]{Feature\\qubits}}
& $N_f=2$ & 39.53 & 0.0045 & 0.9923 \\
& $N_f=3$ & 41.56 & 0.0157 & 0.9739 \\
& $N_f=4$ & 42.59 & \textbf{0.0032} & \textbf{0.9944} \\
& $N_f=5$ & \textbf{37.88} & 0.0202 & 0.9677 \\
& $N_f=6$ & 44.65 & 0.0062 & 0.9895 \\
\midrule
\multirow{4}{*}{\makecell[c]{Circuit\\depth}}
& 10 layers & 39.07 & \textbf{0.0032} & \textbf{0.9945} \\
& 20 layers & 42.59 & \textbf{0.0032} & 0.9944 \\
& 30 layers & \textbf{37.90} & 0.0079 & 0.9863 \\
& 40 layers & 38.73 & 0.0082 & 0.9853 \\
\bottomrule
\end{tabular}
\caption{Quantitative capacity comparison on MNIST. Top: varying feature qubits at $r=20$; bottom: varying circuit depth with $N_f=4$. Best values are bold.}
\label{tab:mnist_qubit_selection}
\label{tab:mnist_depth_selection}
\end{table}
\subsubsection{Effect of Model Capacity}
As shown in Figure~\ref{fig:mnist_nf_ablation}(a), increasing $N_f$ from $2$ to $4$ clearly improves digit shape and edge quality. %
When $N_f=5$, the visual quality becomes close to that of $N_f=4$ while $N_f=6$ shows slight degradation, e.g., parts of the strokes of digit ``8'' in the first row become sticky and connected, and the shape of digit ``9'' is also slightly distorted.
Quantitatively, the fixed circuit depth of $r=20$, $N_f=5$ achieves the lowest FID of $37.88$, but $N_f=4$ gives the best Class JSD and Class Entropy (Table \ref{tab:mnist_qubit_selection}), %
so we adopt $N_f=4$ for the subsequent depth experiments.

Figure~\ref{fig:mnist_nf_ablation}(b) shows 10- and 20-layer circuits perform comparably, while 30-40 layers show mild degradation in both visual quality and class JSD/Entropy. This suggests shallow-to-moderate depth already suffices, with deeper circuits adding optimization difficulty rather than expressivity.

\vspace{0mm}

\begin{figure}[H]
    \centering
    \begin{minipage}[b]{0.5\columnwidth}
        \centering
        \includegraphics[
            width=0.92\linewidth,
            trim=0 203.008pt 0 0,
            clip
        ]{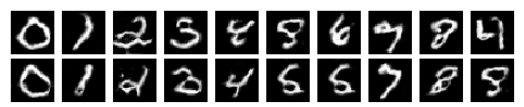}
        \par\vspace{-1mm}
        {\scriptsize $N_f=2$\par}
        \vspace{-0.1mm}
        \includegraphics[
            width=0.92\linewidth,
            trim=0 152.256pt 0 50.752pt,
            clip
        ]{mnist_nf2_nf6_r20_ablation_image_only_representative.pdf}
        \par\vspace{-1mm}
        {\scriptsize $N_f=3$\par}
        \vspace{-0.1mm}
        \includegraphics[
            width=0.92\linewidth,
            trim=0 101.504pt 0 101.504pt,
            clip
        ]{mnist_nf2_nf6_r20_ablation_image_only_representative.pdf}
        \par\vspace{-1mm}
        {\scriptsize $N_f=4$\par}
        \vspace{-0.1mm}
        \includegraphics[
            width=0.92\linewidth,
            trim=0 50.752pt 0 152.256pt,
            clip
        ]{mnist_nf2_nf6_r20_ablation_image_only_representative.pdf}
        \par\vspace{-1mm}
        {\scriptsize $N_f=5$\par}
        \vspace{-0.1mm}
        \includegraphics[
            width=0.92\linewidth,
            trim=0 0 0 203.008pt,
            clip
        ]{mnist_nf2_nf6_r20_ablation_image_only_representative.pdf}
        \par\vspace{-1.45mm}
        {\scriptsize $N_f=6$\par}
        \vspace{-0.2mm}
        {\scriptsize\textbf{(a)} Feature qubits ablation}\par
    \end{minipage}%
    \hfill
    \begin{minipage}[b]{0.5\columnwidth}
        \centering
        \includegraphics[
            width=0.97\linewidth,
            trim=0 152.88pt 0 0,
            clip
        ]{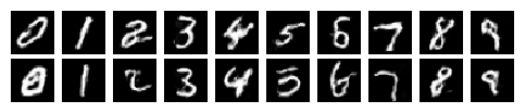}
        \par\vspace{-0.9mm}
        {\scriptsize 10 layers\par}
        \vspace{0.25mm}
        \includegraphics[
            width=0.97\linewidth,
            trim=0 101.92pt 0 50.96pt,
            clip
        ]{mnist_nf4_layer_ablation_image_only_representative.pdf}
        \par\vspace{-0.9mm}
        {\scriptsize 20 layers\par}
        \vspace{0.25mm}
        \includegraphics[
            width=0.97\linewidth,
            trim=0 50.96pt 0 101.92pt,
            clip
        ]{mnist_nf4_layer_ablation_image_only_representative.pdf}
        \par\vspace{-0.9mm}
        {\scriptsize 30 layers\par}
        \vspace{0.25mm}
        \includegraphics[
            width=0.97\linewidth,
            trim=0 0 0 152.88pt,
            clip
        ]{mnist_nf4_layer_ablation_image_only_representative.pdf}
        \par\vspace{-0.9mm}
        {\scriptsize 40 layers\par}
        \vspace{0.25mm}
        {\scriptsize\textbf{(b)} Circuit depth ablation}\par
    \end{minipage}%

    \caption{Representative MNIST samples for two capacity studies. (a) Different numbers of feature qubits at a fixed circuit depth of $r=20$. (b) Different circuit depths using four feature qubits.}
    \label{fig:mnist_nf4_layer_ablation}
    \label{fig:mnist_nf_ablation}
\end{figure}
\endgroup

\section{Conclusion and Discussion}

In this paper, we propose CoQui, an end-to-end GAN framework for image generation based on quantum implicit neural representations, reformulating quantum image generation as a coordinate-conditioned implicit function learning problem. This  decouples image resolution from the number of address qubits and alleviates the inter-pixel  coupling induced by conventional amplitude-to-pixel mapping. We further design a quantum generator with structural inductive biases tailored to this formulation, enhancing model expressiveness. Simulation results show that CoQui achieves better visual quality and quantitative performance than amplitude-mapping-based QGAN baselines while using fewer qubits, and achieves competitive generative performance relative to a classical baseline of comparable parameter scale.
The current evaluation is limited to classical simulation on standard-resolution grayscale benchmarks, without accounting for realistic hardware noise, finite-shot effects, or device connectivity. Future work will extend CoQui to higher-resolution and color image generation while improving scalability, hardware compatibility, and noise robustness toward deployment on real quantum devices.

\bibliography{aaai2027}

\setlength{\leftmargini}{20pt}
\makeatletter\def\@listi{\leftmargin\leftmargini \topsep .5em \parsep .5em \itemsep .5em}
\def\@listii{\leftmargin\leftmarginii \labelwidth\leftmarginii \advance\labelwidth-\labelsep \topsep .4em \parsep .4em \itemsep .4em}
\def\@listiii{\leftmargin\leftmarginiii \labelwidth\leftmarginiii \advance\labelwidth-\labelsep \topsep .4em \parsep .4em \itemsep .4em}\makeatother

\setcounter{secnumdepth}{0}
\renewcommand\thesubsection{\arabic{subsection}}
\renewcommand\labelenumi{\thesubsection.\arabic{enumi}}

\newcounter{checksubsection}
\newcounter{checkitem}[checksubsection]

\newcommand{\checksubsection}[1]{%
  \refstepcounter{checksubsection}%
  \paragraph{\arabic{checksubsection}. #1}%
  \setcounter{checkitem}{0}%
}

\newcommand{\checkitem}{%
  \refstepcounter{checkitem}%
  \item[\arabic{checksubsection}.\arabic{checkitem}.]%
}
\newcommand{\question}[2]{\normalcolor\checkitem #1 #2 \color{blue}}
\newcommand{\ifyespoints}[1]{\makebox[0pt][l]{\hspace{-15pt}\normalcolor #1}}

\clearpage
\onecolumn

\appendix
\renewcommand{\thesection}{Appendix \arabic{section}}

\setcounter{section}{0}
\setcounter{secnumdepth}{1}
\setcounter{figure}{0}
\setcounter{table}{0}

\section{Classical components}
\label{app:classical_components}
The proposed framework contains two classical components: a classical embedding network in the generator and a convolutional critic for adversarial training. The embedding network transforms each spatial coordinate and latent variable into the rotation parameters of the quantum circuit, thereby conditioning the quantum generator on both pixel location and stochastic variation. The critic follows the WGAN-GP formulation and maps an input image to a scalar score through a sequence of strided convolutional layers. 

The classical component of the generator is a classical embedding network. Each pixel coordinate is encoded using a sinusoidal positional embedding with six frequency bands, producing a 26-dimensional coordinate feature. This feature is concatenated with a 10-dimensional latent vector and passed through a three-layer MLP with hidden width 256 and ReLU activations. The MLP outputs 12 rotation angles, corresponding to three rotation angles for each of the four feature qubits. The raw angles are squashed by a $\tanh$ nonlinearity and scaled by $\pi$ before being passed to the quantum circuit.

The discriminator is a classical WGAN-GP critic composed of three stride-2 convolutional layers with $3\times3$ kernels and channel widths $(32,64,128)$, each followed by a LeakyReLU activation with negative slope 0.2. The resulting feature map is flattened and mapped to a scalar critic score by a fully connected layer. All convolutional and dense kernels are initialized with Glorot uniform initialization, and biases are initialized to zero. The detailed architectures and initialization settings of the two components are summarized in Table~\ref{tab:classical_modules}.

\begin{table}[H]
\centering

\resizebox{0.65\columnwidth}{!}{
\begin{tabular}{lll}
\toprule
\textbf{Module} & \textbf{Component} & \textbf{Configuration} \\
\midrule

\multirow{8}{*}{Generator embedding}
& Coordinate input & Two-dimensional pixel coordinate $(x,y)$ \\
& Positional encoding & Sinusoidal encoding with six frequency bands \\
& Coordinate feature dimension & 26 \\
& Latent input dimension & 10 \\
& MLP architecture & Three layers, hidden width 256 \\
& Activation & ReLU \\
& Output & 12 rotation angles for four feature qubits \\
& Output transformation & $\pi \tanh(\cdot)$ \\
\midrule

\multirow{5}{*}{Critic}
& Architecture & Three stride-2 convolutional layers \\
& Kernel size & $3 \times 3$ \\
& Channel widths & 32, 64, and 128 \\
& Activation & LeakyReLU with negative slope 0.2 \\
& Output layer & Flatten followed by a scalar-valued linear layer \\
\midrule

\multirow{2}{*}{Initialization}
& Weights & Glorot uniform initialization \\
& Biases & Zero initialization \\
\bottomrule
\end{tabular}
}
\caption{Architectural details of the classical components used in the final configuration with $N_f = 4$ feature qubits and $r = 20$ re-uploading layers.}

\label{tab:classical_modules}

\end{table}

\section{Training Convergence Analysis}
\label{app:Training Convergence Analysis}

\begin{figure}[h]
    \centering
   \includegraphics[width=0.6\textwidth, trim=0cm 0cm 0cm 0cm, clip]{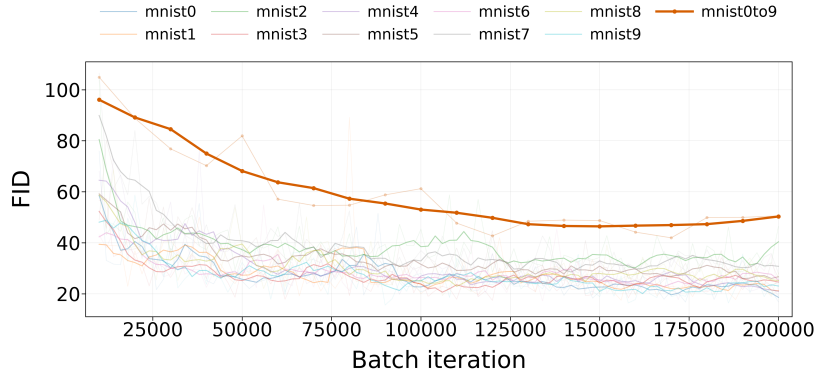}

    \caption{Training FID curves of the jointly trained MNIST 0–9 model and the individual digit-specific models.}
    \label{tab:mnist0to9_highlighted_vs_single_digits}
\end{figure}

To evaluate the generation quality and convergence behavior during training, Figure~\ref{tab:mnist0to9_highlighted_vs_single_digits} presents the FID curves of the jointly trained MNIST 0–9 model and the individual digit-specific models. Since the joint dataset and the single-digit datasets have different target distributions and feature statistics, their FID values are not directly comparable in magnitude. Therefore, we primarily focus on the evolution of FID throughout training rather than its absolute value. To reduce random fluctuations and better illustrate the overall convergence behavior, the curves are smoothed using a moving average with a window size of five evaluation points, while the original FID curves are retained as thinner, lighter background lines to visualize the underlying training fluctuations.
As shown in Figure~\ref{tab:mnist0to9_highlighted_vs_single_digits}, the FID decreases steadily throughout training for both the single-class models and the jointly trained multi-class model, and gradually stabilizes in the later stages. This indicates that the proposed model is able to progressively learn the target data distribution and achieve stable convergence under both single-class and multi-class settings.

\end{document}